\documentstyle[12pt]{article}

\newcommand{\sect}[1]{\setcounter{equation}{0}\section{#1}}

\newcommand{\eq}{\begin{equation}}
\newcommand{\eqa}{\begin{eqnarray}}
\newcommand{\en}{\end{equation}}
\newcommand{\ena}{\end{eqnarray}}
\newcommand{\enn}{\nonnumber \end{equation}}
\newcommand{\spz}{\hspace{0.7cm}}

\def\sk{\vskip .4cm}
\def\noi{\noindent}
\def\om{\omega}
\def\al{\alpha}

\def\Ga{\Gamma}

\def\del{\delta}

\def\epsi{\varepsilon}

\def\part{\partial}

\def\f#1#2{ f^{#1}_{~~#2} }

\def\M#1#2{ M_{#1}^{~#2} }

\def\D{\Delta}

\def\invG{{}_{inv}\Ga}
\def\invX{{}_{inv}\Xi}

\def\dia{{\scriptscriptstyle \Box}}
\def\cvd{\rightline{$\Box\!\Box\!\Box$}\sk}
\def\FF#1#2{O_{#1}{}^{#2} }

\def\DL{\Delta_{\scriptstyle \Ga}}
\def\DR{{}_{\scriptstyle \Ga}\Delta}
\def\DS{\Delta_{\scriptstyle \Xi}}
\def\DD{{}_{\scriptstyle \Xi}\Delta}
\def\N#1#2{N^{#1}{}_{#2}}
\def\vart{\vartheta}
\begin{document}
\begin{titlepage}
\rightline{UCLA/93/TEP/25}
\rightline{November 1993}
\sk
\sk
\sk
\begin{center}{\bf THE SPACE OF VECTOR FIELDS ON QUANTUM GROUPS }\\[6em]
Paolo Aschieri \\[2em]
{\sl Department of Physics \\
University of California, Los Angeles, CA 90024-1547.} \\[6em]
\end{center}
\begin{abstract}
We construct the space of vector fields on quantum groups .
Its elements are products of the known left invariant vector
fields with the elements of the quantum group itself.
We also study the duality between vector fields and 1-forms.
The construction is easily generalized to tensor fields.
A Lie derivative along any (also non left invariant) vector field
is proposed.
These results hold for a generic Hopf algebra.
 \end{abstract}

\vskip 5cm

\noi \hrule
\vskip.2cm
\hbox{\vbox{\hbox{{\small{\it Address after November 1993 :}
Scuola Normale Superiore, 56100 Pisa, Italy. }}}}
\hbox{{\small{\it e-mail:}} ASCHIERI@UX2SNS.SNS.IT}

\end{titlepage}

\newpage
\setcounter{page}{1}

\sect{Introduction}

Following the program of generalizing the differential geometry
structures to the noncommutative case, we construct on a Hopf algebra the
analogue of the space of vector fields.

Indeed in the literature the quantum Lie algebra of left invariant vector
fields as well as the space of 1-forms
has been extensively analized \cite{FRT,Wor,Jurco,Zumino1,AC}
, while the notion of generic vector field on a Hopf
algebra and the duality relation with the space of $1$-forms deserves
more study \cite{Zumino2}.

We will see how left invariant vector fields
generate the whole space of vector fields. This space can be
also characterized as the bicovariant bimodule (vector bundle)
dual to that of 1-forms.

Throughout this paper we will deal  with a Hopf algebra $A$
\cite{Abe,Majid1}
over $\mbox{\boldmath$C$}$with coproduct $\D\::~A\rightarrow A\otimes A
$, counit $\epsi\::~A\rightarrow \mbox{\boldmath$C$}$ and invertible
antipode
$S\::~A \rightarrow A  $.\\
Particular cases of Hopf algebras are quantum  groups, which for
us will be Hopf algebras with one (or more) continuous parameter $q$ ;
when
$q$=$1$ the product ``${\displaystyle \cdot} $'' in $A$ becomes
commutative and
we obtain the algebra of functions on a group{}. When we will speak of
 commutative case we will refer to the Hopf algebra
$C^{\infty}(G)$ of smooth functions on a (compact) Lie group $G$.

In  Section 2 we briefly recall how to associate a
differential calculus to a given Hopf algebra and we emphasize the role
played by the tangent vectors.\\
This construction will be effected along the lines of Woronowicz' work
\cite{Wor}.
Indeed the results in \cite{Wor} apply also to a general Hopf
algebra with invertible antipode (not necessary a compact matrix
pseudogroup). This can be shown by checking that all the formulae
used for the construction described in \cite{Wor} (which are collected in the
appendix  of \cite{Wor})
hold also in the case of a Hopf algebra with invertible antipode
\footnote{Formula (A.22) in \cite{Wor} is the most difficult to prove
and necessitates the further axiom of the invertibility of the
antipode $S$; also the invertibility of the map $s$ in (A.18) relies on
the existence of $S^{-1}$. All the other formulae are direct
consequences of the Hopf algebra axioms.}.

In Section 3 we construct the space of vector fields, while in Section 4
we study the
action of the Hopf algebra on the vector fields; i.e.
we will study the push-forward of vector fields on Hopf algebras.
Then we deal with (covariant and contravariant) tensor fields.

In the last section we propose a Lie derivative and a
contraction operator on differential forms along generic vector fields.
These two operators are basic tools for the formulation of
deformed gravity theories, where the relevant Lie algebra is the
 $q$-Poincar\'{e} Lie algebra.
\sk
\sk
{}
\sect{Differential Geometry on Hopf Algebras}

In the commutative case, given the
differential calculus on a (compact) Lie
group $G$, we can consider the subspace in the space of all
smooth functions $f~:~~G\longrightarrow \mbox{\boldmath$C$}$ defined by:
\eq
R\equiv\{h\in C^{\infty}(G)~ /~~~ h(1_G)=0 \mbox{ and } dh(1_G)=0 \}~,
\label{ideale}
\en
where $1_G $ is the unit of the group.\\
$R$ is a particular ideal of the Hopf algebra $C^{\infty}(G)$.
All the information about the differential calculus  on $G$ is  contained in
$R$.\\
Indeed the space of tangent vectors at the origin of the group is
given by all the linear functionals which annihilate $R$ and any constant
function.
Locally we write a basis as $\{\partial_i|_{1_G}\}$.
Once we have this basis,
using the tangent map (namely
$TL_g$) induced by the left multiplication of the group on itself:
$L_gg'=gg'\;,~\,\forall g,g' \in G$ we can construct a basis of left
invariant vector fields $\{t_i\}$.
Then a generic $1$-form can be written $\rho=f_i\om^i~~[f_i\in
C^{\infty}(G)]$ where $\{\om^i\}$ is the dual basis of
$\{t_i\}$.
Finally, the differential on functions is
\[
d=\om^i t_i ~~~~\mbox{ that is } ~~~~ df=t_i(f)\om^i~.
\]
%Notice that $R=\{h~/~~~h(1_G)=0 \mbox{ and } (dh)(1_G)=0 \}$
%label{Rd}
\sk
In \cite{Wor} the quantum analogue of $R\subset A$ is studied.
Given a Hopf algebra $A$, it turns out that we can always find an $R$ and
construct a differential calculus (in general not  unique).
\sk
The space of tangent vectors on $A$ is then defined as:
\[
T\equiv\{\chi\::~A\rightarrow \mbox{\boldmath$C$}~\:/~~~ \chi ~ \mbox{ linear
functionals, }~\chi (I)=0~
\mbox{ and }~ \chi(R)=0 \} ~,
\]
where $I$ is the unit of $A$ (in the commutative case it is the constant
function \mbox{$I(g)=1$}{} ${}\forall g\in G$).
\sk
Let $\{\chi_i\}~i=1,\,\ldots,n$ be a basis of $T$. Let $x^i\in A$ be $n$
elements such that
\eq
\epsi(x^i)=0~,\label{fine}
\en
\eq
\chi_i(x^j)=\delta^j_i ~.\label{chix}
\en
These $x^i$ can always be found \cite{Wor}.
 We can then define the $n^2$ linear functionals
$\f{i}{j}~:~~A\longrightarrow \mbox{\boldmath$C$}$
\eq
\!\!\!\!\!\!\!\!\!\!\!\!\!\!\!\!\! \forall a \in
A~~~~~~~~\f{i}{j}(a)\equiv\chi_j(x^ia)~.
\en
The $\f{i}{j}$ are well defined [they are independent from the set of $x^i$
chosen to satisfy (\ref{fine}) and (\ref{chix})] and we have
\eq
\chi_i(ab)=\chi_j(a)\f{j}{i}(b) + \epsi(a)\chi_i(b)~.
\label{22bis}
\en
This is the deformed Leibniz rule for the operators $\chi_i$.
In the $q=1$ case, when $R$ becomes the set defined in (\ref{ideale}),
we have $\chi_i=\partial_i|_{1_G} $ ,
 $\f{i}{j}=\delta^i_j\epsi$ and we write (\ref{22bis}) as $
\partial_i(fh)|_{1_G}=(\partial_if|_{1_G})h(1_G) + f(1_G)(\partial_i
h|_{1_G})$.
\sk
For consistency with (\ref{22bis})
the $\f{i}{j}$ must satisfy the conditions:
\begin{eqnarray}
& & \f{i}{j} (ab)= \f{i}{k} (a) \f{k}{j} (b) \label{propf1}\\
& & \f{i}{j} (I) = \del^i_j ~.\label{propf2}
%\\& & (\f{k}{j} \circ S ) \f{j}{i} = \del^k_i ~\epsi~;~~~
%   \f{k}{j} (\f{j}{i} \circ S)  = \del^k_i ~\epsi~.~
%   \label{propf3}
\end{eqnarray}
The space of left invariant  vector fields $\invX$ is easily
constructed from $T$. Using the coproduct $\D$ we define
$\chi*a=(id\otimes \chi)\D(a)$ and
\eq
{}_{inv}\Xi \equiv \{t~ /~~ t=\chi * ~ \mbox{ where } \chi \in T\}
\en
\noi There is a one to one correspondence $\chi_i \leftrightarrow t_i=
\chi_i*$. In order to obtain $\chi_i$ from $\chi_i*$ we simply apply
$\epsi$ (here and in the following we use the notation $\D(a)=a_1\otimes a_2$):
\eq
(\epsi \circ t_i)(a) = \epsi(id \otimes \chi_i)\D(a) =
\epsi(a_1\chi_i(a_2)) = \epsi(a_1)\chi_i(a_2) =
\chi_i(\epsi \otimes id)\D(a) =\chi_i(a)
\en
%\noi(recall (\ref{prop2})

\noi ${}_{inv}\Xi$ is the vector subspace of all linear maps from $A$ to $A$
that
is isomorphic to $T$.

%\noi Moreover $T$ and ${}_{inv}\Xi $ are isomorphic q-Lie algebras where the
%commutator is given by \cite{Wor} \cite{Jurco}:
%\eq
%[\chi,\chi ' ] \equiv (\chi_{(1)} \circ S)*\chi ' *\chi_{(2)}
%\en
%and $\D ' \chi = \chi_{(1)}\otimes\chi_{(2)}$.

We have chosen this perspective to introduce the space of left invariant
vector fields in order to point out that (also in the case of a
general Hopf algebra) it has an existence  on its own,
independent of  the space of 1-forms.
\sk
The space of $1$-forms $\Ga$ is formed by all the elements $\rho$ that
are
 written as formal products and sums of the type
\begin{eqnarray}
\rho=a_i \om^i~. \label{rhoaom}
%\\
%\rho=\om^i b_i \label{rhoomb}
\end{eqnarray}
\noi Here $a_i \in A$ and {$\om^i\,$} $i=1,\,\ldots,n$ is the basis dual to
$\{t_i\}$ .
We express this duality with a bracket:
\eq
\langle\om^i,\chi_j\rangle = \delta^i_j \label{braket} ~~.
\en
Relation (\ref{rhoaom}) tells us that the space of $1$-forms is freely
generated by the elements $\om^i$. By definition any $\rho$ is
decomposed in a unique way as $\rho=a_i\om^i$ and $\Ga$ is a left
$A$-module with the trivial product $b(a_i\om^i)\equiv (ba_i)\om^i$.
$\Ga$ is also a right $A$-module with the following right product:
\begin{eqnarray}
\!\!\!\!\!\!\!\!\!\!\!\!\!\!\!\!\!\!\!
\forall b\in A &~~~~~~~ & \om^i b= (\f{i}{j} * b) \om^j \equiv (id \otimes
\f{i}{j})
\Delta (b)\om^j~. \label{omb}
%\\
%& & a\om^i=\om^j [(\f{i}{j} \circ S^{-1})* a]~. \label{aom}
\end{eqnarray}
{}From this relation it follows that \cite{Wor} :
\begin{eqnarray}
%& & \om^i b= (\f{i}{j} * b) \om^j \equiv (id \otimes \f{i}{j})
%\Delta (b)\om^j \label{omb}\\
\!\!\!\!\!\!\!\!\!\!\!\!\!\!\!\!\!\!\!
\forall a\in A&~~~~~~~ & a\om^i=\om^j [(\f{i}{j} \circ S^{-1})* a] \label{aom}
\end{eqnarray}
and that any $\rho$ can be written in a unique way in the form
\eq
\rho=\om^ib_i \label{rhoomb}
\en
 with $b_i\in A$.
\sk
Finally, the differential operator $~d~:~~A\longrightarrow \Ga~$ can be
defined through the relation:
\eq
\!\!\!\!\!\!\!\!\!\!\!\!\!\!\!\!\!\!\!\!\!\!\!\!\!\!\!\!\!\!\!\!\!
\forall a\in A ~~~~~~~~~~da = (\chi_i * a)\om^i ~.\label{dachi}
\en
As a consequence, the differential calculus obtained has the following
properties:\\
\sk
\noi i) The differential operator satisfies the Leibniz rule
\eq
d(ab)=(da)b+a(db) ~~\forall a,b\in A . \label{Leibniz}
\en
Moreover  any $\rho \in \Ga$ can be expressed as
\eq
\rho=a_{\al} db_{\al} \label{adb}
\en
\noi for some $a_{\al},b_{\al}$ belonging to $A$.\\
The operator $d$ can be extended in a unique way to an exterior
differential $d$  mapping n-forms into (n+1)-forms and such that $d^2=0
$.
\sk
\noi {\sl Remark}. Once we know the operator $d$, the space of tangent
vectors on $A$, like in the commutative case, can be defined as:
\eq
T=\{\chi~ /~~ \chi(a)=0 \mbox{ if and only if } Pda=0\}
\en
where $Pda \equiv S(a_{1})da_{2}$ with
$\Delta(a)=a_{1}\otimes a_{2}.$
The linear map $P$
%$~:~~\Ga\longrightarrow\invG $
is a projection operator; to a given
form $\rho = a_i\om^i$ it associates
the form $P(\rho) = \epsi (a_i)\om^i\,.$
In the commutative case  $\epsi(a_i)$ is the value
that $a_i \in A=C^{\infty}(G)$ takes in the origin $1_G$ of the
Lie group $G$. $P(\rho)$ is then the left invariant 1-form
%that in the ``origin of the not existing
%group'' coincides with the 1-form $\rho$. That is in the commutative case,
%%$Pda$
%is the left-invariant 1-form
whose  value in the origin $1_G$ of the Lie group
equals the value of the 1-form $\rho$ in $1_G$.\\
\sk
\noi ii) The differential calculus is called bicovariant because using $d$ and
the coproduct $\D$ we can define two linear compatible maps $\DL$ and $\DR$
\begin{eqnarray}
& & \DL(adb)=\D(a)(id \otimes d)\D(b),~~~\DL:\Ga\rightarrow A \otimes \Ga
{}~~~{\rm (left~covariance)}
\label{leftco}\\
& & \DR(adb)=\D(a)(d \otimes id)\D(b),~~~\DR:\Ga\rightarrow \Ga \otimes A
{}~~~{\rm (right~covariance)}
\label{rightco}
%& & (id \otimes \DR) \DL = (\DL \otimes id) \DR
%~~~\left({\rm  compatibility between }\right. \DL {\rm and } \DR
%\left.{} \right)
%\label{bicovariance}
\end{eqnarray}
which represent the left and right action of the Hopf algebra on $\Ga$.
In the commutative case they express the
pull-back on 1-forms induced by the left or right multiplication of the
group on itself \cite{AC}.
$\DL$ and $\DR$ are compatible in the sense that $(id \otimes \DR) \DL = (\DL
\otimes id) \DR $.
In the commutative case  this formula tells us that the left and right
actions of the group on $\Ga$ commute: $R^*_g L^*_{g'}=L^*_{g'}R^*_g ~~\forall
\:g,g' \in G.$
{}From the definitions (\ref{leftco}) and (\ref{rightco}) one deduces the
 following properties \cite{Wor}:
\eqa
\DL(a\rho b)=\D (a)\DL(\rho)\D(b),& & ~~~ \DR(a\rho b)=\D(a)\DR(\rho)\D(b)
\label{Dprop0}\\
(\epsi \otimes id) \DL (\rho)=\rho,& & ~~~ (id \otimes \epsi) \DR (\rho)=
\rho \label{Dprop1}\\
(\D \otimes id)\DL=(id\otimes\DL)\DL,& & ~~~ (id\otimes\D)\DR=(\DR\otimes
id)\DR ~.\label{Dprop2}
\ena
An element $\om$ of
$\Ga$ is said to be {\sl left invariant} if
\eq
\DL (\om) = I \otimes \om \label{linvom}
\en
\noi and {\sl right invariant} if
\eq
\DR (\om) = \om \otimes I \label{rinvom}
\en
\sk
We have seen that any $\rho$ is of the form $\rho=a_i\om^i.$ We have
that the $\om^i$ are left invariant and form a basis of $\invG$, the
linear subspace of all left invariant elements of $\Ga$.
Relation (\ref{braket}) tells us that $\invG$ and $\invX$ are dual
vector spaces.\\
\sk
\noi iii) There exists an {\sl adjoint representation} $\M{j}{i}$ of the
Hopf algebra, defined by the right action on the $\om^i$:
\eq
\DR (\om^i) = \om^j \otimes \M{j}{i}~;~~~\M{j}{i} \in A~. \label{adjoint}
\en

The co-structures on the $\M{j}{i}$ can be deduced \cite{Wor}:
\begin{eqnarray}
& & \Delta (\M{j}{i}) = \M{j}{l} \otimes \M{l}{i} \label{copM}\\
& & \epsi (\M{j}{i}) = \delta^i_j \label{couM}\\
& & S (\M{i}{l}) \M{l}{j}=\delta^j_i=\M{i}{l} S (\M{l}{j})
\label{coiM}\end{eqnarray}

The elements $\M{j}{i}$ can be used to
build a right invariant basis of $\Ga$. Indeed the $\eta^i$ defined by
\eq
\eta^i \equiv \om^j S (\M{j}{i}) \label{eta}
\en
are a basis of $\Ga$ (every element of $\Ga$ can be uniquely written
as $\rho = \eta^i b_i$) and their right invariance can be checked
directly .

Moreover, from (\ref{coiM}), using (\ref{eta})
and (\ref{aom})
one can prove the relation
\eq
\M{i}{j} (a * \f{i}{k})=(\f{j}{i} * a) \M{k}{i} \label{propM}
\en
\noi with $a* \f{i}{j} \equiv (\f{i}{k} \otimes id) \D(a)$.
\sk

\sect{Construction of the space of Vector Fields.}

In this section we  study the space $\Xi$ of vector fields over Hopf algebras
defining a
right product between elements of $\invX$ and of $A$.
%$A$-module, (any vector field can be multiplied times a ``function''
%$a\in A$).
\sk
In the commutative case a generic vector field can be written in the
form $f^it_i$ where $\{t_i\}~\,i=1,\,\ldots,n$ is a basis of left invariant
vector fields and $f^i$ are $n$ smooth functions on the group
manifold.

In the commutative case $f^it_i=t_i\dia f^i$ i.e. left and right
products (that we have denoted with $\dia$) are  the same, indeed $ (t_i\dia
f^i)(h)\equiv t_i(h)f^i=f^it_i(h)$.
These considerations lead to the following definition.

Let ${t_i} = {\chi_i *}$ be  a basis in  $\invX$ and
let $a^i$, $\,i=1,\,\ldots,n$ be  generic
elements of $A$:
\sk
\noi {\bf Definition}
\eq
\Xi \equiv \{V~ /~~ V: A \longrightarrow A ~;~ V = t_i\dia a^i\}~,
\label{defff}
\en
where the definition of the right product $\dia$ is given below:
\sk
\noi {\bf Definition}
\eq
\forall a,b \in A  ,\forall t \in {}_{inv}\Xi
{}~~~~~~(t \dia a) b \equiv t(b)a = (\chi * b)a ~.
\en
%We could just have written $ta$ instead of $t\dia a$, but to avoid
%misunderstanding between $t(ab) \in A$ and $t\dia(ab)\in\Xi$ we prefer
%the explicit notation.
The product $\dia$ has a natural generalization to the whole $\Xi$ :
\eq \begin{array}{rcl}
\dia~~:~~& \Xi \times A \longrightarrow & \Xi \nonumber\\
         & ~ (V,a) ~        \longmapsto     &  V\dia a
\end{array}
{}~~~ \mbox{ where }~~~ \forall b \in A~~~ (V \dia a)(b) \equiv V(b)a ~.
\en

\sk
\noi It is easy to prove that $( \Xi ,\dia )$ is a right $A$-module:
\eq
V\dia (a + b) = V\dia a + V\dia b~;~~~
  V\dia (ab) = (V \dia a) \dia b~;~~~
  V\dia(a+b) = V\dia a + V\dia b
\label{amodule}
\en
(we have also $V\dia \lambda a = \lambda V \dia a$ with $\lambda\in
\mbox{\boldmath$C$}\,$).

\noi For example $V\dia (ab) = (V \dia a) \dia b$ because
\[\forall c\in A~~~~ [(V\dia a)\dia b]c = [(V\dia a)(c)]b = (V(c)a)b =
V(c)ab = [V\dia ab]c
.\]
\noi {\sl Note.} \spz  To distinguish the elements $V\dia (ab) \in \Xi $ and
$V(
ab)\in A$ we have not omitted the simbol $\dia$ representing the
right product.
\sk
$\Xi$ is the analogue of the space of derivations on the ring
$C^{\infty}(G)$ of the
smooth functions on the group $G$.
Indeed we have:

\eq V(a+b)=V(a)+ V(b)~~,~~~~~ V(\lambda a) = \lambda
V(a)~~~~\mbox{Linearity}\label{Linearity}
\en
\eq
{}\:~~~V(ab) \equiv (t_i\dia c^i)(ab) = t_j(a)(f^j{}_i*b)c^i + aV(b)
{}~~~~~~\mbox{ Leibniz rule}\label{Leibnizrule}
\en

\noi in the classical case $t_j(a)(f^j{}_i *b)c^i =
V(a)b${}$\,$ (recall $\f{j}{i}=\delta^j_i\epsi\; ;~\:\epsi *b=b).$

\sk
We have seen the duality between $\invG$ and $\invX$. We now extend it to
$\Ga$ and $\Xi$, where $\Ga$ is seen as a left $A$-module (not
necessarily a bimodule) and $\Xi$ is our right $A$-module.

\sk

\noi
{\bf Theorem} 1.
There exists a unique map
$$
\langle~~,~~\rangle ~:~~~\Ga \times \Xi \longrightarrow A
$$
\indent such that:
\sk
\noi 1) $ \forall\: V \in \Xi$; the application
\[
 \langle ~~,V \rangle  ~:~~\Ga \longrightarrow A
\]
%\indent
is a left $A$-module morphism, i.e. is linear and $\langle
a\rho, V\rangle=a\langle\rho, V\rangle$.

\noi 2) $\forall \rho\in \Ga $; the application
\[ \langle \rho,~~\rangle ~:~~\Xi \longrightarrow A\]
is a right $A$-module morphism, i.e. is linear and $\langle
\rho, Vb\rangle=\langle\rho, V\rangle b$.

\noi 3) Given  $\rho\in\Ga$
\eq
\langle\rho,~~\rangle=0
  ~\Rightarrow ~ \rho=0~, \label{Duality1}
\en
\sk
\noi where $\langle\rho,~~\rangle=0$ means
$\langle\rho,V\rangle = 0 ~~\forall V \in \Xi  $.

\noi 4) Given  $V \in \Xi$
\eq
\langle~~,V\rangle=0
{}~\Rightarrow ~ V=0~, \label{Duality2}
\en
\sk
\noi where $\langle~~,V\rangle=0 $ means $\langle\rho,V\rangle = 0 ~~\forall
\rho \in \Ga  .$

\noi 5) On $\invG \times \invX$ the bracket $ \langle~~,~~\rangle$ acts as the
one introduced in the previous section.

\sk
\noi {\sl Remark}. Properties 3) and 4) state that
$\Ga$ and $\Xi$ are dual $A$-moduli, in the sense that
they are dual with respect to $A$.

\sk
\noi {\sl Proof}

\noi Properties 1), 2) and 5) uniquely characterize this map . To prove
the existence of such a map we show that the following bracket
\sk
\noi {\bf Definition}
\eq
\langle\rho,V\rangle \equiv a_{\al}V(b_{\al})~,
\en
 where $a_{\al},b_{\al}$ are elements of $A$ such that
$\rho =a_{\al}db_{\al}$,
%(finite sum over the index $\al$ understood).
satisfies 1),2) and 5).
\sk
\noi We first verify that the above definition is well given, that is:
\[
\mbox{Let } \rho =a_{\al}db_{\al}=a'_{\beta}db'_{\beta}~~\mbox{ then }
{}~~a_{\al}V(b_{\al})= a'_{\beta}V(b'_{\beta})
{}~.\]
\noi Indeed, since \[a_{\al}db_{\al} = a'_{\beta}db'_{\beta}  ~\Leftrightarrow
{}~
       a_{\al}t_i(b_{\al})\om^i = a'_{\beta}t_i(b'_{\beta})\om^i
{}~\Leftrightarrow ~
       a_{\al}t_i(b_{\al}) = a'_{\beta}t_i(b'_{\beta})\]\\
{[we used  the uniqueness of the decomposition (\ref{rhoaom})]}
\noi the definition is consistent
 because $$a_{\al}V(b_{\al})=a'_{\beta}V(b'_{\beta})
{}~\Leftrightarrow ~ a_{\al}t_i(b_{\al})c^i=a'_{\beta}t_i(b'_{\beta})c^i $$
where $V=t_i\dia c^i.$
\sk
\noi Property 1) is trivial since $a\rho =a(a_{\al}db_{\al}) =
(aa_{\al})db_{\al}.$ \\
\noi Property 2) holds since
\[\langle \rho,V\dia c\rangle = a_{\al}(V\dia c)(b_{\al}) =
a_{\al}V(b_{\al})c=\langle \rho,V\rangle c
{}~.\]
%\[ \langle \rho, V+V'\rangle =a_{\al}(V+V')(b_{\al})=a_{\al}V(b_{\al}) +
%a_{\al}V'(b_{\al})= \langle \rho,V\rangle +\langle \rho,V'\rangle ~.\]

\noi Property 5).
Let $\{\om^i\}$ and $\{t_i\}$ be dual bases in $\invG$ and $\invX$.
Since  $\om^i \in
\Ga ~,~ \om^i = a_{\al}db_{\al}$ for some $a_{\al}$ and $b_{\al}$ in $A$.
We can also write $\om^i = a_{\al}db_{\al}= a_{\al}t_k(b_{\al})\om^k$ ,
so that, due to the uniqueness of the decomposition (\ref{rhoaom}),
we have
\[~~~~~~~~~~~~~~~ a_{\al}t_k (b_{\al})=\delta_k^i I~~~~~~~~~~~~(I
\mbox{ unit of} A);\]

\noi we then obtain  \[\langle\om^i,t_j\rangle=a_{\al}t_j(b_{\al})=\delta^i{}_j
I~.\]

\noi Property 3).
Let  $\rho=a_i \om^i \in\Ga~.$ \\
If $\langle\rho , V\rangle =0 ~~\forall V\in \Xi$, in particular
$\langle\rho , t_j\rangle = 0 ~~\forall j=1,...,n$; then
$\mbox{$a_i\langle\om^i , t_j\rangle=0 \Leftrightarrow $} \\ a_j=0$ ,
and therefore $\rho =0~.$
\sk
\noi Property 4).
Let $V = t_i\dia a^i \in \Xi~.$ \\
If $\langle \rho , V\rangle =0 ~~\forall\rho\in\Ga$, in particular
$\langle\om^j  , V\rangle = 0 ~~\forall j=1,...,n$; then
$\mbox{$\langle\om^j , t_i\rangle a^i=0 \Leftrightarrow $} \\ a^j=0$ ,
and therefore $V =0~.$

\cvd
By construction every $V$ is of the form
\[ V=t_i\dia a^i.\]
\indent We can now show the unicity of such a decomposition.
\sk
\noi {\bf Theorem} 2. Any $V \in \Xi$ can be uniquely written in the form
\[V=t_i\dia a^i
\]
\noi {\sl Proof}\\
Let $ V=t_i\dia a^i =t_i\dia a'^i$ then
%$a^i=a'^i ~(i=1,\,\ldots,n)$
\[\!\!\!\!\!\!\!\!\!\!\!\!\!\!\!\!\!\!\!
\forall i=1,\,\ldots,n~~~~~~ a^i=  \langle\om^i , t_j\rangle a^j =
 \langle\om^i , V\rangle = \langle\om^i , t_j\rangle a'^j = a'^i~.
\]
\cvd
\noi Notice that once we know the decomposition of $\rho$ and $V$ in terms of
$\om^i$ and $t_i$, the evaluation of $ \langle~~,~~\rangle$ is trivial:
\[
\langle\rho,V\rangle = \langle a_i\om^i,t_j\dia b^j\rangle = a_i\langle
\om^i,t_j\rangle b^j = a_ib^i~.
\]
Viceversa from the previous theorem $V=t_i\dia\langle\om^i,V\rangle$ and
$\rho = \langle\rho,t_i\rangle \om^i~.$
\sk
\sk
\sk
$\!{}$ We conclude
this section by remarking the three different $\mbox{ways
of looking at $\Xi.$}$

\noi (I)${}~{}~{}$   \spz $\Xi$ as the
set of all deformed derivations over $A$ [see (\ref{defff}),
(\ref{Linearity}) and
(\ref{Leibnizrule})].

\noi (II)${}~{}$  \spz $\Xi$ as the right $A$-module freely  generated by
the {\sl elements} $t_i\,, ~~i=1,\,\ldots,n$. The latter is  the
set of all the {\sl formal} products and sums of the type $t_ia^i$,
where $a^i$ are generic elements of $A$.
Indeed, by virtue of Theorem 2, the map that associates to each
$V=t_i\dia a^i$ in $\Xi$ the corresponding element $t_ia^i $
 is an isomorphism between right $A$-moduli.

\noi (III) $\!$\spz  $\Xi$ as $ \Xi '=
\left\{ U ~:~\Ga \longrightarrow A ,
{}~U \mbox{ linear and } U(a\rho)=aU(\rho) ~\forall a\in A\right\}$, i.e.
$\Xi$ as the dual (with respect to $A$) of the space of 1-forms $\Ga$.
The space $\Xi '$ has a trivial right $A$-module structure: $(Ua)(\rho)
\equiv U(\rho)a~.$
$\Xi$ and $\Xi '$ are isomorphic right $A$-moduli because of property
(\ref{Duality2}) which states that to each $\langle~~,V\rangle \: :~\Ga
\rightarrow A$
 there
corresponds one and only one
$V$.$[\langle~~,V\rangle=\langle~~,V'\rangle\Rightarrow V=V'].$
Every $U\in \Xi '$ is of the form $U = \langle~~, V\rangle$; more precisely,
if $a^i$ is such that
$ U(\om^i)=a^i$ then $U = \langle~~~, t_i\dia a^i\rangle~.$

\sk
\sect{Bicovariant Bimodule Structure.}
\sk
In \cite{Wor} the space of $1$-forms is extensively studied.
A right and a left product are introduced between elements of $\Ga$ and of $A$,
and it is known how to obtain a left product from a right one [e.g.
$\om^ia=(f^i{}_j*a)\om^j$],
i.e. $\Ga$ is a bimodule over $A$.
\sk
Since the actions $\DL$ and $\DR$ are compatible in the sense that:
\eq
(id \otimes \DR) \DL = (\DL \otimes id) \DR
\en
the bimodule $\Ga$ is called a {\sl bicovariant bimodule}. In \cite{Wor} it is
shown that relations
%% FOLLOWING LINE CANNOT BE BROKEN BEFORE 80 CHAR
(\ref{propf1}),(\ref{propf2}),(\ref{aom}),(\ref{propM}),(\ref{copM}),(\ref{couM}) and (\ref{adjoint}) completely characterize the bicovariant
bimodule $\Ga$.
\sk
In Section 3 we have studied the right product $\dia$ and we have seen
that $\Xi$ is a right bimodule over $A$ [see (\ref{amodule})].
In this section we introduce a left product and a left and right action
of the Hopf algebra $A$ on $\Xi$. The left and right actions $\DS$ and
$\DD$ are the $q$-analogue of the push-forward of tensor fields on a
group manifold. Similarly to $\Ga$ also $\Xi$ is a bicovariant bimodule.
\sk
The construction of the left product on $\Xi$, of the right action $\DD$
and of the left action $\DS$ will be effected  along the lines of Woronowicz'
Theorem 2.5 in \cite{Wor}, whose  statement can be explained  in the
following steps:
\sk
\noi {\bf Theorem} 3.
Consider the {\sl symbols} ${t_i}~ (i=1,\,\ldots,n)$
 and let $\Xi$ be the right $A$-module freely generated by them:

\[
\Xi \equiv \{t_i a^i ~/~~a^i \in A\}
\]
Consider functionals $\FF{i}{j}:~A\longrightarrow \mbox{\boldmath$C$}~$
satisfying [see (\ref{propf1}) and (\ref{propf2})]
\begin{eqnarray}
& & \FF{i}{j} (ab)= \FF{i}{k} (a) \FF{k}{j} (b) \label{propF1}\\
& & \FF{i}{j} (I) = \del^i_j~~~~~~~~~~~~~~~. \label{propF2}
\end{eqnarray}
%\\& & (\FF{k}{j} \circ S ) \FF{j}{i} = \del^k_i ~\epsi;~~~
%   \FF{k}{j} (\FF{j}{i} \circ S)  = \del^k_i ~\epsi;~~~
 %  \label{propF3}
\indent Introduce a left product via the definition [see (\ref{aom})]
\sk
\noi {\sl Definition}
%\begin{eqnarray}
%& & \om^i b= (\FF{i}{j} * b) \om^j \equiv (id \otimes \FF{i}{j})
%\Delta (b)\om^j \label{omb}\\
%& & a\om^i=\om^j [(\FF{i}{j} \circ S^{-1})* a] \label{aom}
%\end{eqnarray}
\eq
b(t_ia^i) \equiv t_j{[}(\FF{i}{j}\circ S^{-1})* b{]}a^i~. \label{bta}
\en
It is easy to prove that

\noi i) \spz $\Xi$ is  a bimodule over $A$. (A proof of this first
statement as well as of the following ones is contained in \cite{Wor}).

\rightline{$\Box$}
\sk
Introduce an action (push-forward) of the Hopf algebra $A$ on $\Xi$
\sk
\noi {\sl Definition}
\eq
\DS(t_ia^i)\equiv(I\otimes t_i)\D(a^i)\label{DDta}~.
\en
It follows that

\noi ii) \spz $(\Xi, \DS)$ is a left covariant
 bimodule over $A$, that is
\[
 \DS(a V b)=\D (a)\DS(V)\D(b)~;~~~
 (\epsi \otimes id) \DS (V)=V~;~~~
 (\D \otimes id)\DS=(id\otimes\DS)\DS ~.
\]
\rightline{$\Box$}
\sk
Introduce $\N{i}{j}$ satisfying [see (\ref{propM}),(\ref{copM}) and
(\ref{couM})]
\begin{eqnarray}
& \N{i}{j} (a * \FF{i}{k})=(\FF{j}{i} * a) \N{k}{i} & \label{propN} \\
 & \Delta (\N{j}{i}) = \N{j}{l} \otimes \N{l}{i} &  \label{copN}\\
 & \epsi (\N{j}{i}) = \delta^i_j~~~~~~, &  \label{couN}
\end{eqnarray}
and introduce $\DD $ such that [see (\ref{adjoint})]
\sk
\noi {\sl Definition}
\eq
\DD(a^it_i)\equiv\D(a^i)t_j\otimes \N{j}{i}\label{DS}~.
\en
Then it can be proven that

\noi iii) \spz The elements  [see (\ref{eta})]
\eq
h_i \equiv t_jS(\N{j}{i})\label{heich}
\en
are right invariant: $\DD(h_i) = h_i\otimes I$.
Moreover any $V \in \Xi$ can be expressed in a unique way respectively as
$V=h_ia^i$ and as
$V=b^ih_i$, where $a^i, b^i \in A$.
%the set $\{h_i\}$ forms a basis of $\Xi_{inv}$ where $\Xi_{inv} =\{V
%\in \Xi~/~~ \DD (V) = V\otimes I\} ~.$

\rightline{${}~~~~~~~~~~~~~~~~~~~~~~~~~\Box$}

\noi iv) \spz $ (\Xi, \DD)$ is a right covariant bimodule over $A$,
that is
\[
 \DD(a V b)=\D (a)\DD(V)\D(b)~;~~~ (id \otimes \epsi) \DD (V)=V~;~~~
 (id \otimes \D)\DD =(\DD\otimes id)\DD~.
\]{}
\rightline{$\Box$}

\noi v) \spz The left and right covariant bimodule $(\Xi,\DS,\DD) $ is a
bicovariant
bimodule, that is left and
right actions are compatible:
\[
(id\otimes\DD)\DS=(\DS\otimes id)\DD~.
\]
\cvd

In the previous section we have seen [remark (II)] that the space of vector
fields
$\Xi$ is the free right $A$-module generated by the symbols $t_i$, so
that the above theorem applies to our case.

There are many bimodule structures (i.e. choices of $\FF{i}{j}$) $\Xi$ can
be endowed with.
Using the fact that $\Xi$ is dual to $\Ga$ we request compatibility with
the $\Ga$ bimodule.\\
In the commutative case $\langle f\om^i,t_j\rangle=\langle \om^if,t_j\rangle=
\langle\om^i,ft_j\rangle=\langle\om^i,t_jf\rangle.$\\
In the quantum case we know that $\langle a\om^i,t_j\rangle =
\langle\om^i,t_j\dia a\rangle$ and we require
\eq
\langle\om^i a,t_j\rangle =
 \langle\om^i,at_j\rangle~; \label {wat}\en
this condition uniquely determines the bimodule structure of $\Xi$.
Indeed we have
 \eqa
 \langle\om^i,at_j\rangle &=& \langle\om^i a,t_j\rangle =
\langle (\f{i}{k} *a )\om^k,t_j \rangle = \f{i}{k} *a\langle\om^k,t_j\rangle=
\f{i}{k} * a\,\delta^k_j = \delta^i_l\f{l}{j} * a \nonumber \\
& = &\langle\om^i,t_l \dia
(\f{l}{j} * a)\rangle
\ena
so that
\eq
at_i = t_j \dia (\f{j}{i} * a~). \label{ffS}
\en
We then define
\eq
\FF{i}{j} \equiv \f{j}{i}\circ S
\en
it follows that $\FF{i}{j} \circ S^{-1} = \f{j}{i}$  and
(\ref{ffS}) can be rewritten [see (\ref{aom}) and (\ref{bta})]
\eq
at_j=t_j\dia {[}(\FF{i}{j}\circ S^{-1})*a{]}~.
\en
{\bf Theorem} 4. The functionals $ \FF{i}{j} $ satisfy conditions
(\ref{propF1}) and
(\ref{propF2}).\\
{\sl Proof}
\sk
\noi The first condition $\FF{i}{j}(I)=\delta^j_i $
holds trivially.\\
The second one is also easily checked:
\sk
\noi
\[\begin{array}{rcl}
\!\! \FF{i}{j}(ab) &\!\!\! =\!\!\! & (\f{j}{i}\circ S)(ab) =
\f{j}{i}{[}S(b)S(a){]}
 =\f{j}{k}{[}S(b){]}\f{k}{i}{[}S(a){]}
 =\f{k}{i}{[}S(a){]}\f{j}{k}{[}(S(b){]} \\
               &\!\!\! =\!\!\! &\FF{i}{k}(a) \FF{k}{j}(b)
\end{array}
\]
 {} \cvd
So far $\Xi$ has  a bimodule structure.
$\Xi$ becomes a left covariant bimodule if we define $\DS$ as in
(\ref{DDta}) so that $t_i$ are left invariant vector fields.
% accordingly with (\ref{dachi}).
\sk
As noticed in \cite{Wor} $\M{k}{l}$ and $\f{i}{j}$ are dual in the
sense that $\f{i}{j}(\M{k}{l}) = \Lambda^{il}{}_{kj}$ with
$\Lambda^{il}{}_{kj} =\delta^i_j\delta^l_k $ when
$q = 1$.
This suggests the following definition of the $n^2$
elements $\N{l}{k} \in A$
\sk
\noi {\bf Definition}
\eq
\N{l}{k}=S^{-1}(\M{k}{l})~,
\en
\noi so that also $\N{l}{k}$ and $\FF{j}{i}$ are dual:
 ${}\,\FF{j}{i}(\N{l}{k})=\f{i}{j}(\M{k}{l})=
\Lambda^{il}{}_{kj}$ .
\sk
{\bf Theorem} 5.  The $\N{l}{k}$ elements defined above satisfy relations
(\ref{couN}),
(\ref{copN}) and
(\ref{propN}):
$$
1)~ \epsi (\N{j}{i}) = \delta^i_j
{}~~~2)~ \Delta (\N{j}{i}) = \N{j}{l} \otimes \N{l}{i}
{}~~~3)~ \N{i}{k} (a * \FF{i}{j})=(\FF{k}{i} * a) \N{j}{i}
% & S (\N{i}{l}) \N{l}{j}=\delta^j_i=\N{i}{l} S (\N{l}{j}) &
%\label{coiN}
$$
\sk
\noi {\sl Proof}
\sk
\noi $1)~$ is trivial.
\sk
\noi $2)~~~\!\D(\N{i}{j})=\D{[}S^{-1}(\M{j}{i}){]}=\sigma_A(S^{-1}\otimes
S^{-1})\D(\M{j}{i})=$\\[1em]
${}~~~~~~\sigma_A{[}(S^{-1}(\M{j}{k})\otimes
S^{-1}(\M{k}{i}){]}= \sigma_A(\N{k}{j} \otimes \N{i}{k}) =
\N{i}{k}\otimes\N{k}{j}~$\\
where $\sigma_A$ is the flip in $A\otimes A,~\sigma_A(a\otimes
b)=b\otimes a \mbox{ for any } a,b\in A.$
\sk
\noi $3)~$ We know that [see (\ref{propM})]
\[
\!\!\!\!\!\!\forall\/ a\in A~~~~~~~\M{i}{j} (a * \f{i}{k})=(\f{j}{i} * a)
\M{k}{i}
\]
or equivalently,
\[
{}~~~ ~~~~~~\M{i}{j} [S(a) * \f{i}{k}]=[\f{j}{i} * S(a)] \M{k}{i} ~.
\]
Now
$$\begin{array}{lll}
\!\!\!{[}S(a)*\f{i}{k}{]}&\!\!=\!\!&(\f{i}{k}\otimes
id)\D{[}S(a){]}=(id\otimes\f{i}{k})(S\otimes S)\D(a)=S(id\otimes
\f{i}{k}\circ S)\D(a)\\
&\!\!=\!\!&S(\FF{k}{i}*a)
\end{array}.$$
Similarly,
\[
{[}\f{j}{i}*S(a){]}=S(a*\FF{i}{j})~.
\]
So we can write
\[
S(a*\FF{i}{j})\M{k}{i} = \M{i}{j}S(\FF{k}{i}*a)
\]
for all $a\in A.$
Applying $\:S^{-1}$ to both members of this last expression we obtain
relation \nopagebreak $3).$
 \sk \cvd

Now that we have all the ingredients, the construction of the bicovariant
bimodule
$\Xi$ is easy and straightforward.
For example $\DD$ is given in formula (\ref{DS}).
\sk
We can then conclude that $(\Xi ,\DS ,\/\DD)$ is a bicovariant
bimodule.
\sk
Notice that, since  Theorem 3  completely characterizes a bicovariant
bimodule all the formulas containing the symbols $\f{i}{j}$ or
$\M{k}{l}$ or elements of $\Ga$ are still valid under the
substitutions $\f{i}{j}\rightarrow \FF{i}{j},~~\M{k}{l}
\rightarrow\N{k}{l} $ and $ \Ga\rightarrow \Xi$.
\sk

\sect{Tensor fields}

The construction completed for vector fields is readily generalized to
$p$-times
contravariant tensor fields.

We define $\Xi\otimes\Xi$ to be the space of all elements that
can be written as finite sums of the kind $\sum_i V_i\otimes V'_i $
with $V_i,V'_i\in\Xi$. The tensor product (in the algebra $A{)}$
between $V_i$ and $V'_i$ has the following properties:\\
 ${}~~~~~~V\dia a\otimes
V'=V\otimes aV' ~, a(V\otimes V') =(aV)\otimes V' $ and $ (V\otimes
V')\dia a=V\otimes(V'\dia a) $\\
 so that $\Xi\otimes\Xi $ is naturally a bimodule
over $A$.\\
Left and right actions on $\Xi\otimes\Xi$ are defined by:
\eq
\DS (V \otimes V')\equiv   V_1   {V'} _1 \otimes   V_2 \otimes
  {V'}_2,~~~\DS: \Xi \otimes \Xi \rightarrow A\otimes\Xi\otimes\Xi
\label{DSXiXi}
\en
\eq
\DD (  V \otimes   {V'})\equiv   V_1 \otimes   {V'}_1 \otimes   V_2
  {V'}_2,~~~\DD: \Xi \otimes \Xi \rightarrow \Xi\otimes\Xi\otimes A
\label{DDXiXi}
\en
\noi where as usual $  V_1$, $  V_2$ etc. are defined by
\eq
\DS (  V) =   V_1 \otimes   V_2,~~~  V_1\in A,~  V_2\in \Xi
\en
\eq
\DD (  V) =   V_1 \otimes   V_2,~~~  V_1\in \Xi,~  V_2\in A~.
\en
\noi More generally, we can introduce the action of $\DS$
on $\Xi^{\otimes p}\equiv\underbrace{\Xi \otimes \Xi \otimes \cdots
\otimes \Xi}_{\mbox{$p$-times}}$ as
\[
\DS (  V \otimes   {V'} \otimes \cdots \otimes   {V''})\equiv
  V_1   {V'}_1 \cdots   {V''}_1 \otimes   V_2 \otimes
  {V'}_2\otimes \cdots \otimes   {V''}_2
\]
\eq
\DS~:~~ \Xi^{\otimes p} \longrightarrow
A\otimes\Xi^{\otimes p}~;
\label{DSXiXiXi}
\en
\[
\DD (  V \otimes   {V'} \otimes \cdots \otimes   {V''})\equiv
  V_1 \otimes   {V'}_1 \otimes \cdots \otimes   {V''}_1 \otimes   V_2
  {V'}_2 \cdots   {V''}_2
\]
\eq
\DD~:~~ \Xi^{\otimes p} \longrightarrow
\Xi^{\otimes p}\otimes A~~.
\label{DDXiXiXi}
\en

\noi Left invariance on $\Xi\otimes\Xi$ is naturally defined as
$\DS (  V \otimes   {V'}) = I \otimes   V \otimes   {V'}$ (similar
definition for right invariance), so that for example $t_i \otimes
t_j$ is left invariant, and is in fact a left invariant basis for $\Xi
\otimes \Xi$: each element can be written as $t_i\otimes t_j\dia  a^{ij}$ in a
unique way.

 It is not difficult to show that $\Xi \otimes \Xi$ is a bicovariant bimodule.
In the same way also
$(\Xi^{\otimes p},\DS,\DD)$ is a bicovariant bimodule.
An analogue procedure, using $\DL$ and $\DR$ instead of $\DS$ and $\DD$,
applies also to
$\Ga^{\otimes p}$ the $p$-times tensor product of $1$-forms.
\sk
Any element $v\in \Xi^{\otimes p}$ can be written as
$v=t_{i_1}\otimes\ldots t_{1_p}\dia b^{i_1...i_p}$ in a unique way,
similarly any element $\tau\in\Ga^{\otimes p}$ can be written as
$\tau=a_{i_1...i_p}\om^{i_1}\otimes\ldots \om^{i_p}$ in a unique way.
\sk
It is now possible to generalize the previous bracket
$\langle~~,~~\rangle\,
:\;\Ga\times\Xi\rightarrow A\:$ to $\Ga^{\otimes p}$ and $\Xi^{\otimes p}\;{}$:

\[
\begin{array}{rccll}
\langle~~,~~\rangle ~:~~&\Ga^{\otimes p} \times \Xi^{\otimes p}&
\longrightarrow & A &{} \\
&(\tau,v) &\longmapsto
&\langle\tau,v\rangle &=a_{i_1...1_p}\langle\om^{i_1}\otimes
...\om^{i_p}\, ,\,t_{j_1}\otimes ...t_{j_p}\rangle b^{j_1...j_p}\\
&{}&{}&{}&=a_{{i_1}...i_p}b^{i_p...i_1}
\end{array}
\]
where we have defined
\eq
\langle\om^{i_1}\otimes ...\om^{i_p}\, ,\,t_{j_p}\otimes ...t_{j_1}\rangle
\equiv \langle \om^{i_1},t_{j_1}\rangle ...\langle \om^{i_p},t_{j_p}\rangle
=\delta^{i_1}_{j_1} ...\delta^{i_p}_{j_p} \; ~.
\label{mirrorcoup}
\en
Using definition (\ref{mirrorcoup}) it is easy to prove that
\eq
\langle\tau a,v\rangle=\langle\tau,av\rangle \label{pass}~,
\en
for example
\[\langle\om^{i_1}\otimes\ldots\om^{i_p}a ,
t_{j_1}\otimes\ldots t_{j_p}\rangle=(\f{i_1}{k_1}*\ldots
\f{i_p}{k_p}*a)
\langle\om^{k_1}\otimes\ldots\om^{k_p} , t_{j_1
}\otimes\ldots t_{j_p}\rangle
\]
\[\langle\om^{i_1}\otimes\ldots\om^{i_p} , a
t_{j_1}\otimes\ldots t_{j_p}\rangle=
\langle\om^{i_1}\otimes\ldots\om^{i_p} , t_{l_1
}\otimes\ldots t_{l_p}\rangle
(\f{l_p}{j_p}*\ldots \f{l_1}{j_1}*a)
\]
and these last two expressions are equal if and {\sl only if}
(\ref{mirrorcoup}) holds.

Therefore we have also shown that definition (\ref{mirrorcoup}) is the
only one compatible with property (\ref{pass}), i.e. property
(\ref{pass}) uniquely determines the coupling between $\Xi^{\otimes}$
and $\Ga^{\otimes}.$

It is easy to prove that
the bracket $\langle~~,~~\rangle$ extends
to  $\Ga^{\otimes p}$ and $\Xi^{\otimes p}$ the duality between $\Ga$
and $\Xi$.
\sk
More generally we can define $\Xi^{\otimes}\equiv
A\oplus\Xi\oplus\Xi^{\otimes 2}\oplus\Xi^{\otimes 3}...$
 to be the algebra
of contravariant tensor fields (and
$\Ga^{\otimes}$ that of covariant tensor fields).

The actions $\DS$ and $\DD$ have a natural generalization to
$\Xi^{\otimes}$ so that we can conclude that $(\Xi^{\otimes},\DS,\DD)$
is a bicovariant graded algebra, the graded algebra of tensor fields
over the ring ``of functions on the group'' $A$, with the left and
right ``push-forward'' $\DS$ and $\DD$.

\sect{Lie derivative}
In this last section we propose a definition of Lie derivative along a
generic vector field.
We start with the introduction of  the contraction operator $i_V $ with
$V\in\Xi$.\\
Let $t\in \invX$, the operator $i_t$ on forms is caracterized by:
\[ \begin{array}{l}
\alpha) \hspace{1cm}
	i_{t_{i}}(a)=0 \hspace{.4cm} \; a \in A \\
\beta) \hspace{1cm}
	i_{t_{i}} ( \omega^{j} )= \delta^{j}_{i} I  \\
\gamma) \hspace{1cm}
	i_{t_{i}}
	( \omega^{i_{1} } \wedge \ldots \omega^{i_{n}} )  =
	i_{t_{j}} ( \omega^{i_{1}} ) \f{j}{i} *
	( \omega^{i_{2} } \wedge \ldots \omega^{i_{n}} ) -
	\omega^{i_{1}} \wedge i_{t_{i}}
	( \omega^{i_{2} } \wedge \ldots \omega^{i_{n}} ) \\
\delta) \hspace{1cm}
	i_{t_{i}} ( a \vartheta +\vartheta') =
	a i_{t_{i}} ( \vartheta ) + i_{t_i}(\vartheta')
	 \;\;\;\; \vartheta ,\vartheta'\; \mbox{generic forms} \\
\varepsilon) \hspace{1cm}
	i_{\lambda^{i}t_{i}} =
	\lambda^{i} i_{ t_{i}} \hspace{.4cm}\;\;\;
	\lambda^{i} \in
	\mbox{ \boldmath$C$}
\end{array} \]

These relations uniquely define $i_t$ along $t\in\invX$; its
existence is ensured by the uniqueness of the expansion of a
generic n-form on a basis of left invariant 1-forms :
$\vartheta = a_{ i_{1} i_{2} \ldots i_{n} }
\omega^{i_{1}} \wedge \ldots \omega^{i_{n}}$.
It can be shown that property $\gamma{)}$ holds in the more general case:
\[
\begin{array}{clcl}
                ~~\:  \tilde{\gamma}{)}~~~~ &
i_{t_{i}} ( a_{ i_{1} \ldots i_{n} } \omega^{i_{1} }\wedge
\ldots \omega^{i_{n}} ) &=&
i_{t_{j}} (
a_{ i_{1} \ldots i_{n} }
\omega^{i_{1} } \wedge \ldots \omega^{i_{s}} )
\wedge
\f{j}{i} *
( \omega^{i_{s+1} } \wedge  \ldots \omega^{i_{n}} ) + \\
& & & + (-1)^{s}
a_{ i_{1} \ldots i_{n} }
\omega^{i_{1}} \wedge \ldots \omega^{i_{s}} \wedge
i_{t_{i}} ( \omega^{i_{s+1} } \wedge  \ldots \omega^{i_{n}} )
\end{array}
\]
\noi with $a_{i_1\ldots i_n} \in A$.
\sk
\noi For a generic vector field  $V=t_j\dia b^j$ we define
\sk
\noi {\bf Definition}

%% FOLLOWING LINE CANNOT BE BROKEN BEFORE 80 CHAR
\[\!\!\!\!\!\!\!\!\!\!\!\!\!\!\!\!\!\!\!\!\!\!\!\!\!\!\!\!\!\!\!\!\!\!\!\!\!\!\!\!\!\!\!\!\!\!\!\!\!\!\!\!\!\!\!\!\!\!\!\!\!\!\!
\zeta) \hspace{1cm}
       i_V(\vart)=i_{t_j\dia b^j}(\vart)\equiv i_{t_j} (\vart)b^j ~~~~~~\vart
\mbox{
generic form.}
\]
The following properties are easily proven:
\[ \begin{array}{l}
{}~\alpha') \hspace{1cm}
	i_{V}(a)=0 \hspace{.4cm} \; a \in A \\
{}~\beta') \hspace{1cm}
	i_{V} ( \omega^{j} )= b^j \mbox{ where } V=t_i\dia b^i  \\
\begin{array}{clcl}
    \gamma')~~~~~ &
i_{V} ( a_{ i_{1} \ldots i_{n} } \omega^{i_{1} }\wedge
\ldots \omega^{i_{n}} ) &\!\!\!=\!\!\!&
i_{t_{j}} (
a_{ i_{1} \ldots i_{n} }
\omega^{i_{1} } \wedge \ldots \omega^{i_{s}} )
\wedge
\f{j}{i} *
( \omega^{i_{s+1} } \wedge  \ldots \omega^{i_{n}} )b^i + \\
& &\!\!\! &\!\!\! + (-1)^{s}
a_{ i_{1} \ldots i_{n} }
\omega^{i_{1}} \wedge \ldots \omega^{i_{s}} \wedge
i_{V} ( \omega^{i_{s+1} } \wedge  \ldots \omega^{i_{n}} )
\end{array}\\
\noi\mbox{{ with }} a_{i_1\ldots i_n} \in A.\\
%\gamma') \hspace{1cm}
%	i_{V}
%	(a_{i_1...i_p \omega^{i_{1} } \wedge \ldots \omega^{i_{p}} )  =
%	i_{t_{j}} ( a\omega^{i_{1}} ) \f{j}{i} *
%	( \omega^{i_{2} } \wedge \ldots \omega^{i_{n}} ) -MMMMMMMMMMMMMM
%	\omega^{i_{1}} \wedge i_{t_{i}}
%	( \omega^{i_{2} } \wedge \ldots \omega^{i_{n}} ) \\
{}~\;\delta') \hspace{1cm}
	i_{V} ( a \vartheta +\vartheta') =
	a i_{V} ( \vartheta ) + i_{V}(\vartheta')
	 \;\;\;\; \vartheta ,\vartheta'\;
\mbox{generic forms} \\
{}~\;\varepsilon') \hspace{1cm}
	i_{\lambda V} =
	\lambda i_{V} \hspace{.4cm}\;\;\;
	\lambda \in
	\mbox{ \boldmath$C$}
\end{array} \]
\noi {\sl Remark}. Definition $\zeta{)}$ and property $\delta'{)}$ reduce in
the commutative case to the familiar formulae:
\[
i_{fV}\vart=f\,i_V\vart ~~ \mbox{ and } ~~ i_V(h\vart) = hi_V\vart ~.
\]
\sk
The Lie derivative along left invariant vector fields is given by:
\eq\ell_t(\tau) \equiv (id \otimes \chi)\DR (\tau) = \chi * \tau \;\;
\;\;\;\;\; \ell_t\; : \;\; \Gamma^{\otimes n} \longrightarrow
 \Gamma^{\otimes n}
\;\;\;
\en
where $\chi \in \invX$ is such that $\chi * =t.$
It can be proved that \cite{AC} :

\eq
\ell_{t_i}= i_{t_i}  d +d  i_{t_i}~.
\en
It is then natural to introduce the Lie derivative along a generic
vector field $V$ through the following
\sk
\noi {\bf Definition}
\eq
\ell_{V}= i_{V}  d +d  i_{V}   \label{ellV} ~.
\en
\noi {\bf Theorem} 6. The Lie derivative satisfies the following properties:
\sk
\noi 1) \spz $\ell_Va=V(a)$
\sk
\noi 2) \spz $\ell_Vd\vart=d\ell_V\vart$
\sk
\noi 3) \spz
$\ell_V(\lambda\vart+\vart')=\lambda\ell_V(\vart)+\ell_V(\vart')$
\sk
\noi 4) \spz $\ell_{V\dia
b}(\vart)=(\ell_V\vart)b-(-1)^pi_V(\vart)\wedge db$ \\
\spz where $\vart$ is a generic $p$-form
\sk
\noi 5) \spz \( \ell_V(\rho \wedge\Omega)=\rho\wedge\ell_V(\Omega)+
%% FOLLOWING LINE CANNOT BE BROKEN BEFORE 80 CHAR
\ell_{t_k}(\rho)\wedge(\f{k}{j}*\Omega)b^j+(-1)^pi_{t_k}(\rho)(\f{k}{j}*\Omega)\wedge
db^j\)\\
\spz where $ \rho\in\Ga,~\Omega $ is a left invariant $p$-form and
$V=t_j\dia b^j$.
\sk
\noi {\sl Proof}
\sk
\noi Properties 1), 2), 3) and 4) follow directly from the definition
(\ref{ellV}).\\
Property 5) is also a consequence of definition (\ref{ellV}); the
proof is computational and makes use of the identities
\[ i_{t_j}[(d\rho)\wedge\Omega]=i_{t_k}(d\rho)\wedge\f{k}{j}*\Omega +
d\rho\wedge i_{t_j}\Omega
%[\mbox{ see }\gamma')] ~~~~\mbox{ and }~~~~
{}~~;~~~d(\f{k}{j}*\Omega)=\f{k}{j}*d\Omega~~.
\]
\nopagebreak
\cvd

\sk

\sk
\noi {\bf Acknowledgements.}\\

\noi I would like to thank Leonardo Castellani for fruitful discussions,
for reading the manuscript and for his useful comments.\\
\sk
\noi This work is partially supported
by a grant from Universit\`{a} di Torino, D.R. n.465
del 26.1.1993.

\sk
\sk
\sk
\sk
\sk
\sk

\vfill\eject
\end{document}